\documentclass[8.5pt,twoside,twocolumn]{article}
\usepackage[super,sort&compress,comma]{natbib} 
\usepackage{mhchem}
\usepackage{times,mathptm}
\usepackage{sectsty}
\usepackage{balance} 
\usepackage{color}

\usepackage{graphicx} 
\usepackage{lastpage}
\usepackage[format=plain,justification=raggedright,singlelinecheck=false,font=small,labelfont=bf,labelsep=space]{caption} 
\usepackage{fancyhdr}
\usepackage{epic}
\begin{document}

\thispagestyle{plain}
\fancypagestyle{plain}{
\renewcommand{\headrulewidth}{1pt}}
\renewcommand{\thefootnote}{\fnsymbol{footnote}}
\renewcommand\footnoterule{\vspace*{1pt}%
\hrule width 3.4in height 0.4pt \vspace*{5pt}}

\makeatletter 
\renewcommand\@biblabel[1]{#1}            
\renewcommand\@makefntext[1]%
{\noindent\makebox[0pt][r]{\@thefnmark\,}#1}
\makeatother 
\renewcommand{\figurename}{\small{Fig.}~}
\sectionfont{\large}
\subsectionfont{\normalsize} 

\fancyfoot{}
\fancyhead{}
\renewcommand{\headrulewidth}{1pt} 
\renewcommand{\footrulewidth}{1pt}
\setlength{\arrayrulewidth}{1pt}
\setlength{\columnsep}{6.5mm}
\setlength\bibsep{1pt}

\twocolumn[
  \begin{@twocolumnfalse}
\noindent\LARGE{\textbf{Colloids in liquid crystals: a lattice Boltzmann study}}
\vspace{0.6cm}

\noindent\large{\textbf{J. S. Lintuvuori,$^{\ast}$\textit{$^{a}$} D. Marenduzzo,$^{a}$, K. Stratford,\textit{$^{b}$} and
M. E. Cates\textit{$^{a}$}}}\vspace{0.5cm}

\noindent\textit{\small{\textbf{Received Xth XXXXXXXXXX 20XX, Accepted Xth XXXXXXXXX 20XX\newline
First published on the web Xth XXXXXXXXXX 200X}}}

\noindent \textbf{\small{DOI: 10.1039/b000000x}}
\vspace{0.6cm}

\noindent \normalsize{
We propose a hybrid lattice Boltzmann algorithm to simulate the hydrodynamics of colloidal particles inside a liquid crystalline host. {{To validate our algorithm, we}} study the static and the microrheology of a colloid in a nematic, with tangential anchoring of the director field at the particle surface, and we confirm theories and experiments showing that the drag force in a nematic is markedly anisotropic. We then {{apply our method to}} consider the case of a colloid inside a cholesteric, and with normal anchoring at the surface. We show that by tuning the ratio between particle size and cholesteric pitch it is possible to control the defect configuration around the particle, and to stabilise {{novel}} figure-of-eight or highly twisted loops close to the colloid surface.}
\vspace{0.5cm}
 \end{@twocolumnfalse}
  ]

\section{Introduction}


\footnotetext{\textit{$^{a}$~School of Physics and Astronomy, University of Edinburgh, Mayfield Road, Edinburgh EH9 3JZ, UK}}
\footnotetext{\textit{$^{b}$~EPCC, University of Edinburgh, Mayfield Road, Edinburgh  EH9 3JZ, UK. }}


A suspension of colloidal particles in liquid crystals is an intriguing example of soft material, which has been receiving increasing attention in the past few years among physicists and chemists~\cite{Poulin,Araki,Stark,Terentjev,Fukuda}. Liquid crystals are anisotropic materials whose average coarse grained orientation may be described by a vector field, known as the {\it director field}. The presence of colloidal particles creates a disturbance in the director alignment, which results in the formation of disclination lines or other topological defects. The interactions between two or more colloids embedded in a liquid crystal is governed by the elasticity of the surrounding medium and by defect-defect interactions, both of which are long range. Experiments, theory and simulations have shown that these elastic and topological long range interactions may provide in practice a route to the formation of a striking variety of self-assembled spatially organised structures, making colloids in liquid crystals a promising example of a material with controlled and programmable macroscopic properties.

For instance, Poulin et al.~\cite{Poulin} have shown that large colloidal particles in nematics with homeotropic (i.e. normal) anchoring of the director field at their surface form lines or sometimes branched networks. More recently, Araki et al.~\cite{Araki} have performed simulations of smaller particles, again with normal anchoring and in a nematic host, and found that the particles form very long-lived aggregates, which are sometimes quasi-planar and are held together by a single disclination loop embracing all the particles in a cluster. Several theoretical works based on liquid crystal elasticity~\cite{Stark,Terentjev,Fukuda} have successfully explained why particles of different sizes behave so differently: this is due to the different defect configuration minimising the liquid crystalline free energy close to the colloid surface. For small particle size, there is a disclination at the equatorial plane of the colloid, known as a ``Saturn ring'', which has quadrupolar symmetry. {{The disclination line has topological charge -1/2, whereas the total charge of the loop is -1}}. For large size, instead, the minimal energy configuration corresponds to a point defect of integer topological charge (equal to -1) on one side of the particle, so that the resulting colloid plus defect system now has a dipolar symmetry. The transition between Saturn ring and dipole may also be triggered by increasing the strength of the anchoring at a fixed particle size.

Previous simulation work on the topic has focussed on this case of spherical particles in nematics with normal anchoring~\cite{Araki,Fukuda,SoftMatterMiha,Miha1,Miha2,Miha3}. With the exception of the previously referred simulations in \cite{Araki} which solve for the hydrodynamics of very viscous and rigid droplets, the majority of other numerical work have considered a purely elastic model in which the nematodynamics of the liquid crystalline host is neglected. This method provides a way to energy minimisation and as such as has been very useful to characterise dimerisation and oligomerisation in nematic colloids, as well as the formation of entangled disclination networks in regular two or three-dimensional arrays of particles. On the other hand, the non-hydrodynamic framework by necessity rules out e.g. rheological studies, which are experimentally useful ways to characterise soft materials macroscopically. Furthermore, it is possible that the self-assembly of a large number of colloids may be dynamically controlled, in which case the prediction of the final structure by theory requires a faithful description of the host hydrodynamics. 

Our programme in this work is therefore to take a different route to study the hydrodynamics of colloidal particle in liquid crystals, by means of hybrid lattice Boltzmann (LB) simulations. LB has long proved to be a powerful method to study colloidal suspensions in Newtonian or binary fluids~\cite{Ladd94_1,Ladd94_2,Ladd,Ignacio,Kevin}, and we show here that it promises to be equally effective to probe the hydrodynamics of colloids in liquid crystals. {{Our algorithm is the first hybrid LB based algorithm to study the hydrodynamics of colloidal particles in liquid crystals. The use of an LB formalism to solve the Navier-Stokes equation allows us to take flow effects into considerations, unlike most of the previous work in the literature (with the exception of the different algorithm used in~\cite{Araki} which however approximates the solid particles with viscous fluid droplets).}}

{{To validate our LB method, we first}} reproduce earlier experimental and theoretical observations that the drag force on a sphere moving in a nematic liquid crystal is anisotropic. {{In the second part of our work, we instead apply our algorithm to find new results on the problem of a colloidal particle~\footnote{Note that we here use the term ``colloid'' also to refer to a single particle.} with homeotropic anchoring immersed in a cholesteric liquid crystal. In particular, we describe the disclination pattern and the director field surrounding a colloidal particle with homeotropic anchoring in a cholesteric host.}} We show that by varying the ratio between particle size and cholesteric pitch it is possible to control the disclinations around the particle -- upon increasing size this changes from a Saturn ring to a figure-of-eight to a twisted loop which covers the surface more uniformly (note that with the parameters we chose the range of particle sizes is such that in a nematic the Saturn ring is always the stable configuration for particle with homeotropic (normal) surface anchoring). Previous related but distinct work had either focussed on colloidal particles in twisted nematics~\cite{Miha3}, realised via conflicting anchoring at the boundary planes and a nematic liquid crystals, or in blue phases~\cite{Miha5}, where the emphasis was on exploring the interactions between nanoparticles and the disclination lines making up the blue phase lattice. Our results instead are relevant to the standard cholesteric phase, and our predictions could be tested with microscopy experiments in the bulk.

The study of colloidal intrusions in a cholesteric host requires the accurate resolution of an additional length scale, the cholesteric pitch. In practice, LB simulations of cholesterics need several (in our case 16 or more) lattice sites to resolve the cholesteric half pitch. As we are interested in the case in which the particle size can compete and be larger than the cholesteric pitch (this is likely to be the experimentally relevant case), the colloid needs to be rather large in size. This immediately leads to numerical limitations as the Brownian time scales with the cube of particle size, hence problems in which we need colloids to diffuse significant distances are computationally costly. These can only be performed via parallel simulations in supercomputers. While we are exploring these studies, here we limit ourselves to one-particle simulations, which do not require such long equilibration times yet are interesting as they lead to new physics and can already answer a number of interesting questions.



\section{Simulation method}

\subsection{Liquid crystalline equations of motion}

The thermodynamics of the cholesteric solvent is determined by the Landau - de Gennes free energy ${\cal F}$, or equivalently its density ${f}$, which is expressed in terms of a (traceless and symmetric) tensorial order parameter~\cite{beris} $\mathbf{Q}$. Here we briefly review the theory, and refer the reader to e.g. \cite{beris,Henrich09,Henrich10} for more details. The free energy density we use is
\begin{align}
{f} & = \tfrac{A_0}{2} \bigl( 1 - \tfrac{\gamma}{3} \bigr) Q_{\alpha \beta}^2 
           - \tfrac{A_0 \gamma}{3} Q_{\alpha \beta}Q_{\beta \gamma}Q_{\gamma \alpha}
           + \tfrac {A_0 \gamma}{4} (Q_{\alpha \beta}^2)^2 \notag \\ 
	 & \quad + \tfrac{K}{2}\bigl( \nabla_{\beta}Q_{\alpha \beta}\bigr)^2
	   + \tfrac{K}{2} 
           \bigl( \epsilon_{\alpha \gamma \delta} \nabla_{\gamma} Q_{\delta \beta} 
           + 2q_0 Q_{\alpha \beta} \bigr)^2. 
\label{eq:FreeEnergy}
\end{align}
In the formula above, $A_0$ is a constant, $K$ is an elastic constant, $q_0=2\pi/p$, $p$ being the cholesteric pitch, and $\gamma$ represents a temperature-like control parameter for thermotropic liquid crystals. In our notation Greek indices denote Cartesian components and summation over repeated indices is implied. Note the limit $q_0=0$ corresponds to a nematic host. The distortion free energy in that limit is equivalent to the more commonly used $\frac{K}{2}\left(\nabla_{\alpha} Q_{\beta \gamma}\right)^2$ up to a total divergence. 

In the Beris-Edwards model, the evolution of the $\mathbf{Q}$ tensor is described by the following equation~\cite{beris}
\begin{equation}
D_t \mathbf{Q} 
= \Gamma  \Bigl( \tfrac{-\delta {\cal F}}{\delta \mathbf{Q}} + \tfrac{1}{3}\, 
\text{Tr} \Bigl( \tfrac{\delta {\cal F}}{\delta \mathbf{Q}} \Bigr) \mathbf{I} \Bigr)  .
\label{eqQevol}
\end{equation} 
Here $\Gamma$ is a collective rotational diffusion constant, $\tfrac{\delta {\cal F}}{\delta \mathbf{Q}}$ is a functional derivative and $D_t$ is the material derivative for rod-like molecules~\cite{beris}, while ${\rm Tr}$ denotes the tensorial trace. The term in brackets is the molecular field, $\mathbf{H}$, which ensures that in the absence of flow $\mathbf{Q}$ evolves towards a minimum of the free energy. The velocity field, $\mathbf{u}$, obeys the continuity equation and the Navier-Stokes equation,
\begin{equation}
\rho \left(\partial_t+u_{\beta}\nabla_{\beta}\right) u_{\alpha}=\nabla_{\beta}P_{\alpha\beta}+\eta \nabla_{\beta}\left(\nabla_{\alpha}u_{\beta}+\nabla_{\beta}u_{\alpha}\right),
\end{equation}
where $\rho$ is the fluid density and $\eta$ is an isotropic viscosity. The stress tensor appropriate for liquid crystal hydrodynamics is explicitly given by:
\begin{eqnarray}
P_{\alpha\beta}= &-&P_0 \delta_{\alpha \beta} +2\xi
(Q_{\alpha\beta}+\tfrac{1}{3}\delta_{\alpha\beta})Q_{\gamma\epsilon}
H_{\gamma\epsilon}\\\nonumber
&-&\xi H_{\alpha\gamma}(Q_{\gamma\beta}+\tfrac{1}{3}
\delta_{\gamma\beta})-\xi (Q_{\alpha\gamma}+\tfrac{1}{3}
\delta_{\alpha\gamma})H_{\gamma\beta}\\ \nonumber
&-&\nabla_\alpha Q_{\gamma\nu} \frac{\delta
{\cal F}}{\delta\nabla_\beta Q_{\gamma\nu}}
+Q_{\alpha \gamma} H_{\gamma \beta} -H_{\alpha
 \gamma}Q_{\gamma \beta} ,
\label{BEstress}
\end{eqnarray}
where $\xi$ is related to the aspect ratio of the molecules and $P_0$ is an isotropic pressure \cite{beris}. 

There are a few dimensionless parameters which are useful to keep in mind and which determine the physics of the liquid crystal host and of the colloidal particle. On one hand, the equilibrium thermodynamics of chiral liquid crystals is mainly controlled by two quantities, the chirality $\kappa$ and the reduced temperature $\tau$, which are given in terms of previously defined quantities as:
\begin{eqnarray}
\kappa & = & \sqrt{{108 K q_0^2}/{A_0 \gamma}} \\ \nonumber
\tau & = & 27 (1-\gamma/3)/\gamma.
\end{eqnarray}
The physics of a colloidal particle moving in the liquid crystal is on the other hand primarily controlled by the Ericksen number,
\begin{equation}
{\rm Er}  = \frac{\gamma_1 vR}{K},~\mathrm{with}~\gamma_1=\frac{2q^2}{\Gamma}
\end{equation}
which measures the ratio between viscous and elastic forces.  In the expression for ${\rm Er}$, $\gamma_1$ is the rotational viscosity of the liquid crystal, $q$ is the degree of ordering in the system (for uniaxial case with director $\hat{\bf n}$, $Q_{\alpha\beta}=q(n_\alpha n_\beta-\delta_{\alpha\beta}/3)$) and $v$ and $R$ are the velocity and radius of the colloidal particle respectively. 

The Navier-Stokes and order parameter Beris-Edwards equations characterising the liquid crystalline host can be solved via a 3D hybrid lattice Boltzmann/finite difference (LB/FD) algorithm detailed elsewhere~\cite{CatesSoftMatter}.

In the following simulations we have modelled a liquid crystal with these parameters in simulation units: $A_0=1.0,~K \simeq 0.065,~\xi=0.7,~\gamma = 3.0$, $q=1/2$ and $\Gamma =0.5$. They lead to effective temperature $\tau=0$, and rotational viscosity $\gamma_1=1$. Assuming an elastic constant of $6.5$pN~\footnote{Within the one elastic constant theory used here, the Frank elastic constants are given in terms of $K$ as $2Kq^2$.}, rotational viscosity of 1 Poise (and a colloidal diameter of $1\mu m$), we can map the simulations units for force, time and velocity onto $\sim 100$ pN, $1 \mu s$ and $0.03 \mu$m/s, respectively. We note that these parameters are close to those of liquid crystals typically used in experiments or devices, such as 5CB.

\subsection{BBL and hybrid dynamics}

To represent colloids as moving solid objects within the LB method, we
have made use of the well-established procedure of bounce-back on links (BBL)
\cite{Ladd94_1,Ladd94_2,Ladd}.
In the hybrid LB/FD approach used here, BBL is retained for
the LB density distributions, which are simply reflected at the solid-fluid
surface with a correction which depends on the local surface
velocity. The resulting change in momentum is
summed over the links to give the net hydrodynamic
force on the colloid, which is then used to update the
particle velocity, and
hence position, in a molecular dynamics-like step. Boundary
conditions for the finite difference equations for the order
parameter tensor are dealt with in a different way.

First, we note that the assignment of solid and fluid lattice
nodes for the order parameter follows that for the density:
inside and outside are distinguished using the nominal radius
of the colloid $a_0$ and its position. It is useful, in addition,
to think about a series of control volumes surrounding each lattice
node whose faces are aligned with the lattice (Fig. 1). A set of
these faces constitute the solid-fluid boundary in the hybrid picture.

Boundary conditions for $Q_{\alpha\beta}$ are of two types:
homeotropic, where the director  $n_\alpha^0$ is aligned with
the local unit normal to the surface $\hat{s}_\alpha$, and planar,
where the director lies in the plane of the tangent to the
surface. The exact orientation in the planar case
 is determined by a projection of the existing director 
onto the local tangent plane, with unit normal $\hat{\bf s}$. 
(In this case the director is allowed to relax in the finite difference step).
For either choice of director at the surface, we
may set the corresponding value of $Q_{\alpha\beta}$ at
lattice nodes immediately inside the surface via
\begin{equation}
Q_{\alpha\beta}^0 = S^0 (n_\alpha^0 n_\beta^0
- {\scriptstyle\frac{1}{3}}\delta_{\alpha\beta})
\end{equation}
where the constant $S^0$ controls the degree of surface order.
This allows us
to compute, at all fluid nodes, the derivatives
$\nabla_\gamma Q_{\alpha\beta}$ and $\nabla^2 Q_{\alpha\beta}$
using the same finite difference stencil. This allows the
molecular field and hence the diffusive terms in the Beris
Edwards equation to be computed.

Also appearing in the Beris-Edwards equations is the velocity
gradient tensor (this comes in the material derivative $D_t$ in
Eq.~\ref{eqQevol}), which can be handled in a similar fashion
close to the colloid. The velocity field at solid nodes
immediately inside the colloid surface are set to the solid
body velocity ${\bf u}=\bf{v} + \bf{\omega} \times \bf{r}$ (${\bf \omega}$ 
denotes the angular velocity of the colloid). Again,
the velocity gradient tensor $\partial_\alpha u_\beta$ may
be computed using the same stencil at all fluid nodes.

Advective fluxes of order parameter are computed at the faces
of the control volumes, and the boundary condition is zero
normal flux at solid-fluid interfaces. At the given time step, the discrete
colloid is effectively stationary, so the no normal flux condition assumes a stationary boundary.

The force on the fluid originating from the order parameter
is computed via the discrete divergence of the stress
$P_{\alpha\beta}$. In the fluid, this is implemented by
interpolating $P_{\alpha\beta}$ to the control volume faces
and taking differences between faces in each direction. This
method has the advantage that, with the introduction of colloids,
an interpolation/extrapolation of $P_{\alpha\beta}$ to the
solid-fluid boundary is possible. This allows one to compute the
divergence of the stress at fluid nodes adjacent to the colloid. 
At the same time, the discrete equivalent of
\begin{equation}
F_\alpha^\mathrm{coll} = \int P_{\alpha\beta} \hat{s}_\beta dS
\end{equation}
is found
by summing $P_{\alpha\beta}$ over the relevant solid-fluid control
volume faces. By construction, this ensures that momentum lost by
the fluid is gained by the colloid, i.e., global momentum is
conserved.

Finally, movement of the colloid across the lattice is accompanied
by changes in its discrete shape. The events necessitate the
removal or replacement of fluid information. For the replacement
of fluid at newly exposed lattice nodes, this means an
interpolation of nearby order parameter values in the fluid to
provide the new information. This is analogous to what is done
for the LB distributions.

\begin{figure}[h]
\centering
\includegraphics{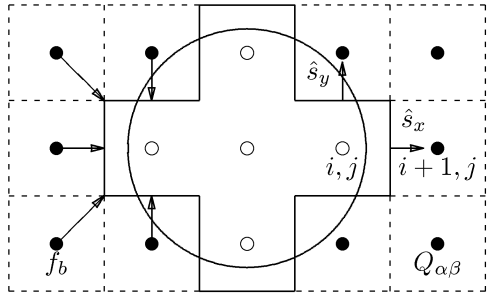}
\caption{The colloid (represented by the solid circle) moves continuously
across the lattice. Lattice sites inside are designated solid, and those
outside fluid (open and closed points, respectively). In the lattice
Boltzmann picture (left) the surface is defined by a set of links
$f_b$, which involve discrete vectors $\mathbf{c}_b \Delta t$ which
connect fluid and solid sites. For the order parameter (right), the
colloid is represented by the set of faces, e.g., that between sites
$i,j$ and $i+1,j$ with unit normal $\hat{s}_x$. Discretisation effects
are found to be negligible for radii greater than about 5 lattice units.}
\end{figure}

\section{Results}
\subsection{A colloidal particle in a nematic liquid crystal: planar anchoring}

{{To first validate our algorithm,}} we start from the case of a colloidal particle in a nematic liquid crystal. We consider planar anchoring, {{which has been well studied theoretically and experimentally before, but, due to the degeneracy in the allowed director configurations, is a more challenging case for simulations with respect to the normal anchoring. Indeed most simulation work has focussed on the latter}}~\cite{Miha1,Miha2,Miha3,Miha4} (but see Ref.~\cite{SoftMatterMiha} for an interesting exception using mixed planar and normal anchoring on a single ``Janus'' particle). 

\begin{figure}[h]
\centering
  \includegraphics[width=\columnwidth]{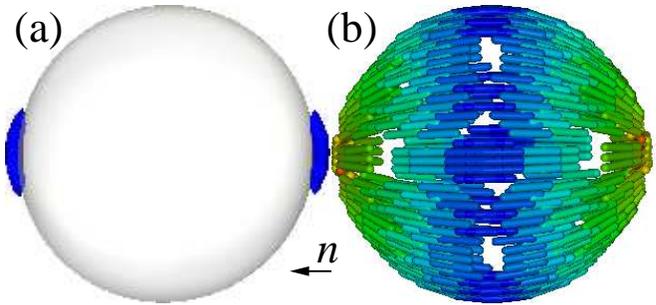}
  \caption{(a) A colloid with planar surface anchoring in nematic host results in a defect structure with two surface defects of topological charge $+1$ (boojums). (b) 3D plot of the director field close to the particle surface with color coding indicating the alignment with the far field nematic director, $\hat{\bf n}$ (marked by the arrow). Blue and red denote parallel and perpendicular alignment respectively. Frame (b) was produced with QMGA.~\cite{qmga}}
  \label{fgr:disclination}
\end{figure}
A colloidal particle with planar surface alignment inside a nematic liquid crystal leads to {{a defect structure confined to the particle surface, with two surface point defects on opposite sides of the particle along the far field nematic director (Fig.~\ref{fgr:disclination}(a)\footnote{Note that the point defects appear as patches in our rendering as we plot isosurfaces of the order parameter. The defects are then the small volumes enclosed by these isosurfaces, within which the order parameter is below a given threshold, which in our case was fixed to 0.435.}) Each of the defects has integer topological charge (equal to +1, in agreement with the Poincare theorem, see e.g.~\cite{lavrentovich}).} Fig.~\ref{fgr:disclination}(b) shows the director field (individual directors are represented by spherocylinders) close to the partice surface. Here the color coding quantifies the alignment between the local and the far field nematic director (marked by the arrow): blue indicates parallel alignment and red perpendicular alignment. 

The bounce-back on links allow us to have approximately no slip on the particle surface. This is very well achieved everywhere apart from close to the boojums, where the maximum spurious velocity is about $0.001$ (in LB units) after 50,000 steps.

\begin{figure}[h]
\centering
  \includegraphics[width=\columnwidth]{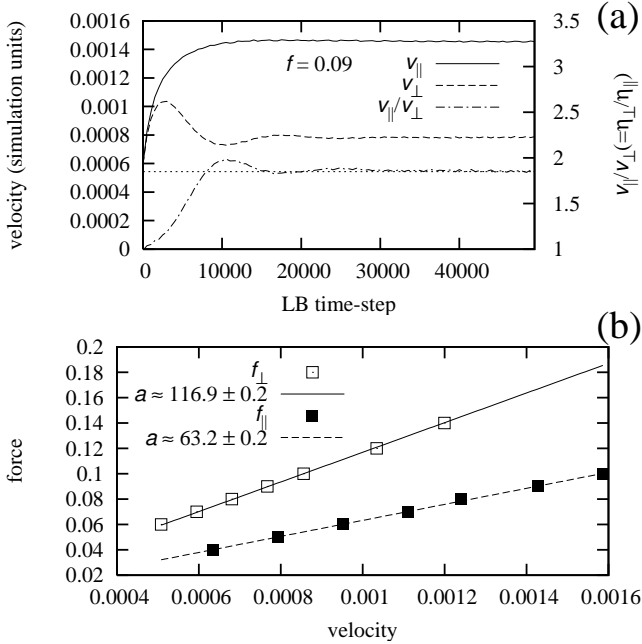}
  \caption{(a) Plot of the velocity versus time for a colloidal particle pulled along (solid line) and perpendicular to (dashed line) the nematic director with constant force $f=0.09$ (a) (left axis). In the same panel the ratio between the two velocities is shown (dot-dashed line, right axis). The dotted line is the ratio $a_{\perp}/a_{\parallel}=\eta_{\perp}/\eta_{\parallel}$, obtained from the fits in (b) (see the text for details). (b) Plot of the force--velocity curve for the system pulled perpendicular (open symbols) and along (closed symbols) the nematic director. The fits are performed via Eq.~(\ref{eq:force}) in the text.(Particle radius is $R=16$, and the isotropic viscosity $\eta$, which enters the Navier-Stokes equation, is equal to 0.1 in simulation units.)}
  \label{fgr:viscosity}
\end{figure}
One of the main strength of our LB algorithm is that, on top of recovering the correct statical properties of the colloid plus director field, it readily allows hydrodynamic studies. We exploit this by directly simulating an active microrheological experiment, in which a colloid is dragged inside a nematic, both perpendicularly to and along the far field nematic director field (again the director is forced to lie tangentially to the particle surface). This is interesting in view of the passive microrheology experiments and drag calculations for this system, which both predict markedly anisotropic behaviours and effective viscosities in the two cases~\cite{Stark2,Poulin2}.   
Fig.~\ref{fgr:viscosity}a shows a typical time series of the instantaneous velocity of a colloid pulled with constant force (in this case $f=0.09$) along, and perpendicular to, the far field director. Fig.~\ref{fgr:viscosity}b shows the corresponding velocity--force curves. These force--velocity data were fitted by the means of the following formula for the drag force,
\begin{equation}\label{eq:force}
f=av.
\end{equation}
Here $a$ has the dimension of viscosity, which is different for perpendicular and parallel dragging. In the former case, $a_{\perp}\approx 116.9 \pm 0.2$, whereas in the latter $a_{\parallel}\approx 63.2 \pm 0.2$. Assuming an effective Stokes' law to hold for the viscous drag, such that $f=6\pi\eta R$, we can now interpret our results as pointing to the existence of two distinct effective microrheological viscosities $\eta^{\mathrm{eff}}$, which can be probed when dragging a spherical particle inside a nematic liquid crystal. We find that the drag force is anisotropic with the ratio between these two effective microviscosities perpendicular/parallel to the nematic director $\eta^{\mathrm{eff}}_{\perp}/\eta^{\mathrm{eff}}_{\parallel}=a_{\perp}/a_{\parallel}\approx 1.850\pm 0.004$. This result compare quite well to $\eta^{\mathrm{eff}}_{\perp}/\eta^{\mathrm{eff}}_{\parallel}\approx 1.6$ and $\eta^{\mathrm{eff}}_{\perp}/\eta^{\mathrm{eff}}_{\parallel}\approx 1.72$ observed earlier by experiments~\cite{Poulin2} and predicted theoretically~\cite{Stark2} for a Stokes drag experienced by a Saturn-ring droplet in nematic liquid crystal at low Er (for simulations with larger Er see~\cite{Araki2,dePablo}). However our results hold for planar anchoring in which the defect structure is a pair of boojums (Fig. 2) rather than a Saturn ring. This points to a rather weak dependence of the viscosity ratio on the details of the anchoring of the director fields at the particle surface.

\subsection{A colloidal particle in a cholesteric: normal anchoring}

As mentioned in the Introduction, by varying particle size and/or anchoring strength, one can trigger a switch in the configuration of the defect close to a particle with normal anchoring immersed in a nematic liquid crystal, from a Saturn ring to a dipolar defect. {{As discussed several times in the literature on colloids in liquid crystals~\cite{Poulin,Miha3,Smalyukh05,dePablo}, the defect structure will typically}} determine the interparticle interactions, and hence the ultimate self-assembled structure, in a system at finite colloid density. Hence it is interesting to ask whether there may be other ways to control defect topology by altering other properties in a colloidal suspension in a liquid crystal. To explore this possibility, we have studied the statics of spherical particle of various sizes inside, this time, a cholesteric liquid crystal. This case is readily treatable within our framework, by just selecting the Landau-de Gennes distortion free energy appropriate for cholesterics. 

Fig.~\ref{fgr:chiral_disc} shows the minimal energy configurations found by our algorithm. We focus on a range of particle sizes for which the free energy inside a nematic host is minimised through a Saturn ring arrangement (Fig.~\ref{fgr:chiral_disc}a). We have then considered a cholesteric liquid crystal, with pitch, $p=32$, lattice units (giving chirality $\kappa=0.3$), and various particle sizes. For a particle with $R=p/4$ (so that its diameter is half the pitch), the Saturn ring elongates and twists up into a figure-of-eight structure. This is similar to a metastable structure found in a colloidal dimer in nematics~\cite{Araki}: in this case however this state is stable and just requires a single colloid. Increasing the size of the particle, the disclination loop twists and becomes longer, providing a more uniform coverage of the particle surface (Figs.~\ref{fgr:chiral_disc}c,d). 

\begin{figure}[h]
\centering
  \includegraphics[width=\columnwidth]{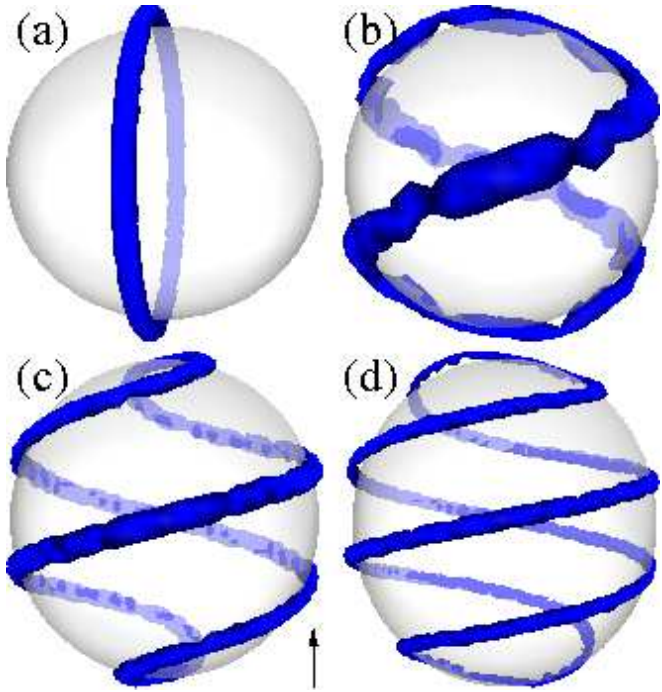}
  \caption{Defect structure close to the particle surface. In the nematic limit ($R/p=0$) a Saturn ring defect is formed (a). For cholesteric case ($R/p>0$) twisted Saturn rings are formed: (b) $R/p=1/4$, (c) $R/p=1/2$ and (d) $R/p=3/4$. (The arrow denotes the direction of the cholesteric helix.)}
  \label{fgr:chiral_disc}
\end{figure}

{{In order to better compare with existing and future experiments, it is also useful to plot the cholesteric layers close to the particle, as these may be observed with optical microscopy experiments. These are shown in Fig. 5, where we have color-coded one of the components of the director field in the plane perpendicular to the direction of the cholesteric helix (see caption of Fig. 5). It can be seen that there is a significant bending of the layers right above and below the colloidal particle, whereas laterally there is a shrinking and dilation of the cholesteric layers. In particular, the layers in which the director ordering conflicts with that imposed by the anchoring shrink, and this is more marked close to the disclination line. What is perhaps surprising is the limited distortion of the liquid crystal, which extends to less than its size.}}

\begin{figure}
\centering
  \includegraphics[width=\columnwidth]{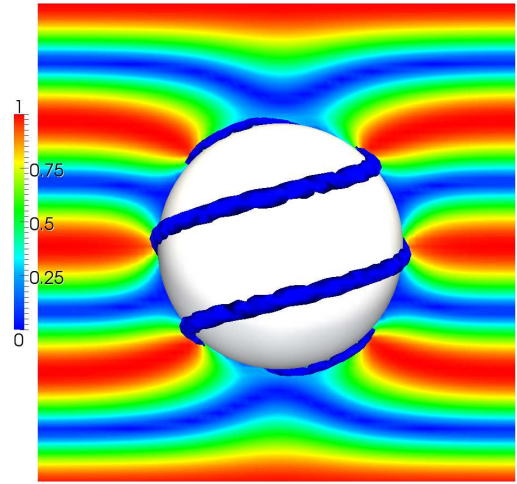}
  \caption{{{Cut of the director field in the $yz$ plane, $\hat{z}$ being the direction of the cholesteric helix, see coordinate axes in the picture. The director field is color-coded according to its projection on $\hat{x}$ (see color bar). The disclination lines are also shown in blue. The radius of the colloidal particle with is equal to $p/2$.}}}
  \label{fgr:chiral_layers}
\end{figure}

{{Interestingly, in a recent experimental work performed at the Kent State University~\cite{Senyuk08}\footnote{See also the information available online at the Kent State web-site at http://www.lci.kent.edu/Lavrentovich/Overview.html$\#$rotation.}, a colloidal particle was analysed while it was sedimenting. It was found that the particle rotated while translating. More relevant to our analysis, the defect structure formed around the particle was determined by optical microscopy, and it was found that it consisted in spirals going around the particles, akin to the ones we found with our LB algorithm. Furthermore, via fluorescence confocal polarizing microscopy the deformation of the cholesteric layers was determined, and the results are in good qualitative agreement with our snapshot in Fig. 5.  It should also be noted that our picture refers to a steady state result, for a particle which is at rest in equilibrium -- it can be imagined that a particle in motion such as a sedimenting one may result in larger deformations.}}

{{We close this section with an overview of possible future calculations which our results may open up. In order to understand the self-assembly potential of colloidal particles in liquid crystals,}} it would be important to probe the different effective interactions that are caused by the defect configurations which we have described. As mentioned in the Introduction, in a nematic liquid crystal Saturn ring forming colloids aggregate in clusters kept together by disclination loops. It might well be that the twisted chiral Saturn rings found here in cholesterics aggregate or interact quite differently. Finally, it would also be interesting to simulate the micro and macrorheology of these particles, both in isolation and in concentrated solutions. Our LB algorithm is in principle capable of addressing these issues, and we hope to explore some of them elsewhere.


\section{Conclusions}

In conclusion, we have {{presented a new hybrid}} lattice Boltzmann algorithm to simulate the hydrodynamics of colloidal particles in a liquid crystal. 

{{To validate our algorithm, we have first}} considered the case of a colloid in a nematic, with tangential anchoring at the director field at the particle surface. We have reproduced the known defect structure, consisting of two surface defects {{(boojums)}} of integer topological charge, and also addressed the hydrodynamic problem of the viscous force which is felt by one such particle as is dragged inside the liquid crystal. As expected from theoretical calculations and experiments focussing on Saturn rings, we have found that the drag force is anisotropic, with the colloid moving more easily along the far field director field than perpendicular to it. The direct counterpart of our simulations would be microrheology experiments, e.g. performed with optical tweezers in the fixed-force mode, where the force felt by the particle is maintained constant via a feedback mechanism.

We have then considered the static of a spherical particle embedded with normal anchoring in a cholesteric liquid crystal, and found that by varying the particle size to cholesteric pitch ratio it is possible to control the defect configuration close to the surface of the particle. In particular, the Saturn ring which embraces the particle in the nematic limit twists up and elongates, to form e.g. a figure-of-eight loop when the particle size equals the cholesteric half pitch. Furher increase in particle size leads to longer and more twisted loops, which more uniformly cover the surface. Therefore the defect configuration may be controlled by tuning the size to pitch ratio, and it would be interesting to understand what consequences this tunable topology has on the self-assembly of a solution of such colloidal particles.

We acknowledge funding by EPSRC Grants EP/E030173 and EP/E045316. MEC is funded by the Royal Society.




\footnotesize{
\bibliography{rsc} 
\bibliographystyle{rsc} 
}

\end{document}